\documentclass[12pt]{article}

\usepackage{amsmath}
\usepackage{amssymb}
\usepackage[dvips]{graphicx}
\usepackage{cite}

\makeatletter
 
  \@addtoreset{equation}{section}
 \makeatother

\newcommand {\n}{\nonumber \\}
\newcommand {\tr}{\mbox{tr}}

\begin{document}
\setlength{\oddsidemargin}{0cm}
\setlength{\baselineskip}{7mm}

\begin{titlepage}
\begin{normalsize}
\begin{flushright}
\begin{tabular}{l}
February 2009
\end{tabular}
\end{flushright}
  \end{normalsize}

~~\\

\vspace*{0cm}
    \begin{Large}
       \begin{center}
         {Covariant Formulation of M-Theory II}
       \end{center}
    \end{Large}
\vspace{1cm}

\begin{center}
           Matsuo S{\sc ato}\footnote
           {
e-mail address : msato@cc.hirosaki-u.ac.jp}\\
      \vspace{1cm}
       
         {\it Department of Natural Science, Faculty of Education, Hirosaki University\\ 
 Bunkyo-cho 1, Hirosaki, Aomori 036-8560, Japan}

\end{center}

\hspace{5cm}

\begin{abstract}
\noindent

We propose a supersymmetric model that defines M-theory. It possesses SO(1, 10) super Poincare symmetry and is constructed based on the Lorentzian 3-algebra associated with U(N) Lie algebra. This model is a supersymmetric generalization of the model in arXiv:0902.1333. From our model, we derive BFSS matrix theory and IIB matrix model in the naive large N limit by taking appropriate BPS vacua.

\end{abstract}

\vfill
\end{titlepage}
\vfil\eject

\setcounter{footnote}{0}

\section{Introduction}
\setcounter{equation}{0}

M-theory is strongly believed to define string theory non-perturbatively. Although BFSS matrix theory \cite{BFSS} and IIB matrix model \cite{IKKT} describe some non-perturbative aspects of string theory, covariant dynamics have not been derived, such as the covariant membrane action or a longitudinal momentum transfer of D0 branes. This is because $SO(1, 10)$ symmetry is not manifest in these models. 

BFSS matrix theory and IIB matrix model can be obtained by the matrix regularization of the Poisson brackets of the light-cone membrane theory \cite{deWHN} and of Green-Schwarz string theory in Schild gauge \cite{IKKT}, respectively. Because the regularization replaces a two-dimensional integral over a world volume by a trace over matrices, BFSS matrix theory and IIB matrix model are one-dimensional and zero-dimensional field theories, respectively. On the other hand, the bosonic part of the membrane action can be written covariantly in terms of Nambu bracket as $T_{M2} \int d^3 \sigma \sqrt{ \{ X^L, X^M, X^N \}^2}$, where L, M, N run $0, 1, \cdots, 10$ \cite{Nambu}\footnote
{
The authors of \cite{LeePark} proposed an action of the supermenbrane whose derivatives appear only through Nambu brackets. At this time, we do not know the way to regularize it because it contains higher than second order terms.
}. One way to obtain a SO(1, 10) manifest field theory is to regularize the Nambu bracket $\{ X^L, X^M, X^N \}$ \cite{Yoneya} by 3-algebra $[X^L, X^M, X^N]$ \cite{BLG1,Gustavsson,BLG2}\footnote
{
A formulation of M-theory by a cubic matrix action was proposed by Smolin \cite{Smolin1, Smolin2, Azuma} }. In this case, we obtain a zero-dimensional field theory. In order to reproduce BFSS matrix theory and IIB matrix model, the 3-algebra needs to associate with ordinary Lie algebra. Recently, the authors of \cite{Lorentz0, Lorentz1, Lorentz2, Lorentz3, Iso} found that such 3-algebra needs to possess a metric with an indefinite signature. 

In this letter, we propose a supersymmetric model that defines M-theory:
\begin{equation}
S=-\frac{1}{12}<[X^L, X^M, X^N]^2>
+ \frac{1}{4}<\bar{\Psi} \Gamma_{MN}[X^M, X^N, \Psi]> \label{intro}, 
\end{equation}
which is a supersymmetric generalization of the bosonic action in our previous paper \cite{bosonicM}\footnote
{We discussed a relation between (\ref{intro}) and the matrix models briefly in the appendix of \cite{bosonicM}. 
}. The bosons $X^L$ and the Majorana fermions $\Psi$ are spanned by elements of Lorentzian 3-algebra associated with U(N) Lie algebra. This action defines a zero-dimensional field theory and possesses manifest SO(1, 10) symmetry. By expanding fields around appropriate BPS vacua, we derive BFSS matrix theory and IIB matrix model in the naive large N limit.

\vspace{1cm}

\section{Action of M-theory}
\setcounter{equation}{0}

We propose a following model that defines M-theory,
\begin{equation}
S=-\frac{1}{12}<[X^L, X^M, X^N]^2>
+ \frac{1}{4}<\bar{\Psi} \Gamma_{MN}[X^M, X^N, \Psi]>, \label{generalaction}
\end{equation}
where $X^L$ with $L = 0, 1, \cdots, 10$ are vectors and $\Psi$ are Majorana spinors of SO(1, 10). This action defines a zero-dimensional field theory and possesses manifest SO(1,10) symmetry. There is no coupling constant.

 $X^M$ and $\Psi$ are spanned by elements of the Lorentzian 3-algebra associated with U(N) Lie algebra,
\begin{eqnarray}
X^M&=&X^M_{-1} T^{-1} +X^M_0 T^0 +X^M_i T^i, \n
\Psi&=&\Psi_{-1} T^{-1} +\Psi_0 T^0 +\Psi_i T^i, \label{3-algelements}
\end{eqnarray}
where $i=1, 2, \cdots, N^2.$
The algebra is defined by 
\begin{eqnarray}
&&[T^{-1}, T^a, T^b]=0, \n
&&[T^0, T^i, T^j]=[T^i, T^j]=f^{ij}_{\quad k} T^k, \n
&&[T^i, T^j, T^k] = f^{ijk} T^{-1},
\end{eqnarray}
where $a,b=-1,0,1,2, \cdots, N^2$ and $f^{ijk} = f^{ij}_{\quad l} h^{lk}$ is totally anti-symmetrized. $[T^i, T^j]$ is a Lie bracket of the U(N) Lie algebra. 
The metric of the elements is defined by
\begin{eqnarray}
&& <T^{-1}, T^{-1}>=0, \quad <T^{-1}, T^{0}>=-1, \quad <T^{-1}, T^{i}>=0, \n
&& <T^{0}, T^{0}>=0, \quad <T^{0}, T^{i}>=0, \quad <T^{i}, T^{j}>=h^{ij}.
\end{eqnarray}

By using these relations, the action is rewritten as 
\begin{eqnarray}
S&=& \tr (-\frac{1}{4}(X_0^L)^2 [X_M, X_N]^2 
+\frac{1}{2} (X_{0}^{M}[X_M, X_N])^2 \n
&&\quad+\frac{1}{2}X^M_0\bar{\Psi}\Gamma_{MN}[X^N, \Psi]
-\frac{1}{2}\bar{\Psi}_0\Gamma_{MN}\Psi[X^M, X^N]),\label{rewritten}
\end{eqnarray}
where $X^M=X^M_i T^i$ and $\Psi=\Psi_i T^i$. There should be no ghost in the theory, because $X_{-1}^M$ or $\Psi_{-1}$ do not appear in the action\footnote
{
Ghost-free Lorentzian 3-algebra theories were studied in \cite{ghost1, ghost2}.
}.

Let us summarize symmetry of the action. First, gauge symmetry is $N^2$-dimensional translation $\times$ U(N) symmetry associated with the Lorentzian 3-algebra \cite{Lorentz1}. Second, there are two kinds of shift symmetry. First one is eleven-dimensional translation symmetry generated by
\begin{equation}
\delta X^M = c^M, \label{translation}
\end{equation}
where $X^M \in U(N)$, $c^M \in U(1)$ and the other fields are not transformed. Second one is a part of supersymmetry generated by
\begin{equation}
\delta_1\Psi=\epsilon_1, \label{super0}
\end{equation}  
where $\Psi \in U(N)$, $\epsilon_1 \in U(1)$ and the other fields are not transformed.

Third, the action is invariant under another part of supersymmetry transformation,
\begin{eqnarray}
\delta_2 X^M &=& i \bar{\epsilon_2} \Gamma^M \Psi \label{super1}\\
\delta_2 X^M_0 &=& i \bar{\epsilon_2} \Gamma^M \Psi_0 \label{super2}\\
\delta_2 \Psi &=& - \frac{i}{2} [X^L, X^M] X_{0}^{N} \Gamma_{LMN} \epsilon_2 \label{super3}\\
(\delta_2 \bar{\Psi}_0)\tilde{\Psi} &=& -\delta_0 S, \label{super4}
\end{eqnarray}
where $\tilde{\Psi}=\frac{1}{2} \tr(\Gamma_{MN} \Psi [X^M, X^N])$ and $\delta_0 S$ is the variation of the action (\ref{rewritten}) under (\ref{super1}), (\ref{super2}) and (\ref{super3}).

We should note that the above super transformation is slightly different with a straightforward analogue to that of the BLG theory for multiple M2-branes, which is given by\begin{eqnarray}
\delta X^M &=& i \bar{\epsilon} \Gamma^M \Psi \n
\delta \Psi &=& - \frac{i}{6} [X_L, X_M, X_N] \Gamma^{LMN} \epsilon \label{covarianttrans}.
\end{eqnarray}
If we decompose this transformation, (\ref{super1}), (\ref{super2}) and (\ref{super3}) are the same, but (\ref{super4}) is different. In the analogue case, $\delta \Psi_0 = 0.$ There is no such symmetry because $\delta_0 S \neq 0$. However, in our Lorentzian case the action does possess supersymmetry because $\delta_2\Psi_0$ cancels $\delta_0 S$.

$\delta_1$ and $\delta_2$ form supersymmetry in eleven dimensions because the commutator of the super transformations of $X^M$ results in the eleven-dimensional translation (\ref{translation}) as, 
\begin{equation}
(\delta_1 \delta_2 - \delta_2 \delta_1)X^M = -i \bar{\epsilon_1} \Gamma^M \epsilon_2,
\end{equation}
where $X^M \in U(N)$. 
Therefore, eigen values of $X^M \in U(N)$ should be interpreted as eleven-dimensional space-time\footnote
{This kind of mechanism and interpretation was originally found in \cite{IKKT}. 
}. In the next section, we will derive BFSS matrix theory and IIB matrix model in the large N limit from our model. In this derivation, $X^i (i=1, \cdots, 9) \in U(N)$ and $X^{i} (i=0, \cdots 9) \in U(N)$ are identified with matrices in BFSS matrix theory and IIB matrix model respectively. Therefore, our interpretation is consistent with the space-time interpretation in these models.  

\vspace{1cm}

\section{BFSS Matrix Theory and IIB Matrix Model from M-Theory}
\setcounter{equation}{0}
Our theory possesses a large BPS moduli that includes simultaneously diagonalizable configurations. Such configurations should be independent vacua in the large N limit because of the supersymmetry. By treating appropriate BPS configurations as backgrounds, we derive BFSS matrix theory and IIB matrix model in the large N limit. 
  
We consider backgrounds
\begin{eqnarray}
\bar{X}^{\mu}&=&p^{\mu}=\mbox{diag}(p^{\mu}_1, p^{\mu}_2, \cdots, p^{\mu}_N), 
\label{bg1} \\
\bar{X}^I &=& 0 \label{bg4} \\
\bar{X}_0^M&=&\frac{1}{g} \delta^M_{10},
\label{bg2} \\
\bar{\Psi}&=&\bar{\Psi}_0=0 \label{bg3}
\end{eqnarray}
where $\mu=0, 1, \cdots, d-1 (d \le 10)$ and $I = d, \cdots, 10$.
$(p_{0}^i, p_{1}^i, \cdots, p_{d-1}^i)$ ($i=1, \cdots, N$) represent N points randomly distributed in a d-dimensional space. There are infinitely many such configurations. $X_0^M$ represents an eleven-dimensional constant vector. By using SO(1,10) symmetry, we can choose (\ref{bg2}) as a background without loss of generality. $g$ will be identified with a coupling constant. $g \to \infty$ corresponds to $\bar{X}_0^M=0$, which leads to SO(1,10) symmetric vacua. These configurations are BPS states, namely (\ref{bg1}), (\ref{bg4}), (\ref{bg2}) and (\ref{bg3}) satisfy $\delta_2 \Psi= \delta_2 \Psi_0=0$.

 Because of the supersymmetry, all the backgrounds (\ref{bg1}),  (\ref{bg4}), (\ref{bg2}) and (\ref{bg3}) should be treated as independent vacua and fixed in the large N limit \cite{KS}, as in the discussion of Higgs mechanism. Thus, we do not integrate $X_0^M$, $\Psi_0$ or the diagonal elements of $a_{\mu}$ and we expand the fields around the backgrounds as, 
\begin{eqnarray}
X_{\mu}&=&p_{\mu}+a_{\mu}. \n
X_I&=&x_I \n
\Psi&=&\psi,
\label{expansion}
\end{eqnarray} 
where we impose a chirality condition 
\begin{equation}
\Gamma^{10}\psi=\psi \label{chiral}. 
\end{equation}

Under these conditions, the first term of the action (\ref{rewritten}) is rewritten as
\begin{eqnarray}
S_1&=&\tr (-\frac{1}{4}(X_0^L)^2 [X_M, X_N]^2) \n
   &=&-\frac{1}{4g^2}\tr([p_{\mu}+a_{\mu}, p_{\nu}+a_{\nu}]^2 
+ 2[p_{\mu}+a_{\mu}, x^I]^2 
+ [x^I, x^J]^2).  
\end{eqnarray}
The second term is  
\begin{eqnarray}
S_2&=&\frac{1}{2} \tr((X_{0}^{M}[X_M, X_N])^2) \n
   &=&\frac{1}{2g^2} \tr([p_{\mu}+a_{\mu}, x^{10}]^2 + [x^{10}, x^I]^2).  
\end{eqnarray}
As a result, the total action is independent of $x^{10}$ as follows,
\begin{eqnarray}
S&=&-\frac{1}{g^2}\tr(\frac{1}{4}[p_{\mu}+a_{\mu}, p_{\nu}+a_{\nu}]^2 
+ \frac{1}{2}[p_{\mu}+a_{\mu}, x^i]^2 
+ \frac{1}{4}[x^i, x^j]^2 \n
&& \qquad + \frac{g}{2}\bar{\psi}\Gamma^{\mu}[p_{\mu}+a_{\mu}, \psi] 
+ \frac{g}{2}\bar{\psi}\Gamma^{i}[x_i, \psi]), \label{lattice}
\end{eqnarray}
where $i,j = d, \cdots, 9$.
In the large N limit, this action is equivalent to
\begin{eqnarray}
S&=&-\frac{1}{g^2} \int d^d \sigma \tr( \frac{1}{4}F_{\mu\nu}^2 
- \frac{1}{2}(D_{\mu} x^i)^2 
+ \frac{1}{4}[x^i, x^j]^2 \n
&& \qquad \qquad \qquad
+ \frac{i}{2} \bar{\psi} \Gamma^{\mu} D_{\mu} \psi 
+ \frac{1}{2} \bar{\psi} \Gamma^i [x_i, \psi]),
\end{eqnarray}
where $\psi$ is redefined to $\frac{1}{\sqrt{g}}\psi$.
This fact is proved perturbatively and non-perturbatively in the large N limit as in the case of the large N reduced model \cite{EK,Parisi,BHN,GK}.

Under the conditions (\ref{bg1}) - (\ref{chiral}), the super transformations (\ref{super1}) and (\ref{super3}) reduces to
\begin{eqnarray}
\delta a^{\mu} &=& i \bar{\epsilon} \Gamma^{\mu} \psi \n
\delta x^{I} &=& i \bar{\epsilon} \Gamma^I \psi \n
\delta \psi &=& -\frac{i}{2g}
([p_{\mu}+a_{\mu}, p_{\nu}+a_{\nu}] \Gamma^{\mu\nu}
+[x^i, x^j]\Gamma^{ij}) \epsilon, 
\end{eqnarray}
by which (\ref{lattice}) is invariant. Moreover, (\ref{super2}) and (\ref{super4}) reduces to
\begin{eqnarray}
\delta X^M_0 &=& 0 \n
\delta \Psi_0 &=& 0,
\end{eqnarray}
because the action (\ref{rewritten}) reduces to the action (\ref{lattice}) and $\delta_0 S=0$. This is consistent with the fact that $X^M_0$ and $\Psi_0$ are fixed.

Therefore, if we choose the BPS backgrounds with $d=1$, we obtain 
BFSS matrix theory in the large N limit,
\begin{equation}
S=\frac{1}{4g^2} \int d \tau \tr(
2(D_{0} x^i)^2 
- [x^i, x^j]^2
- \frac{i}{2} \bar{\psi} \Gamma^{0} D_{0} \psi 
- \frac{1}{2} \bar{\psi} \Gamma^i [x_i, \psi]). 
\end{equation}
If we choose those with $d=0$, we obtain
IIB matrix model in the large N limit,
\begin{equation}
S=-\frac{1}{4g^2}\tr(
[x^i, x^j]^2
+ \frac{1}{2} \bar{\psi} \Gamma^i [x_i, \psi]). 
\end{equation}
We also obtain matrix string theory \cite{Motl, BS, DVV} when $d=2$ and $AdS_5/CFT_4$ \cite{Malda} when $d=4$.

\vspace{1cm}

\section{Conclusion and Discussion}
\setcounter{equation}{0}

In this letter, we have proposed a covariant action that defines M-theory. It possesses SO(1, 10) super Poincare symmetry. In this model, the eleven-dimensional space-time is given by eigen values of the U(N) part of the bosonic fields $X^M$. From this action, by choosing appropriate BPS vacua we have derived BFSS matrix theory and IIB matrix model in the large N limit. By using these relations, we can directly discuss covariant dynamics that have not been derived, such as covariant membrane and M5-brane actions and a longitudinal momentum transfer of D0 particles.

We impose the chirality condition (\ref{chiral}) by hand in the process to derive the matrix models. Therefore, we also need to check that the chirality condition (\ref{chiral}) is automatically satisfied in the large N limit. A planer diagram that includes fermions with opposite chirality may be forbidden. Supersymmetry may play a crucial role.

As a first step, we have shown that our theory reproduces large N dynamics of the critical matrix models. In this limit, only planer diagrams contribute. As a second step, we should examine whether our theory includes all dynamics of the matrix models in order to check that it is a complete action of M-theory.

\end{document}